\documentclass[prl,reprint, superscriptaddress, preprintnumbers, nofootinbib]{revtex4-2}
\usepackage[ngerman,english]{babel}
\usepackage{booktabs}
\usepackage{lineno,hyperref}
\usepackage{lipsum}
\usepackage{graphicx}
\usepackage{dcolumn}
\usepackage{amsmath}
\usepackage{amsfonts} 
\graphicspath{{}{}{.}}
\modulolinenumbers[5] 
\begin{document}
\title{Improved $S$-factor of the $^{12}$C(p,$\gamma$)$^{13}$N reaction at $E\,=\,$320-620~keV and the 422~keV resonance}

\author{J. Skowronski}\affiliation{Universit\`a degli Studi di Padova, 35131 Padova, Italy} \affiliation{INFN, Sezione di Padova, 35131 Padova, Italy}

\author{E. Masha}\affiliation{Helmholtz-Zentrum Dresden-Rossendorf, 01328 Dresden, Germany}

\author{D. Piatti}\email{denise.piatti@pd.infn.it}\affiliation{Universit\`a degli Studi di Padova, 35131 Padova, Italy} \affiliation{INFN, Sezione di Padova, 35131 Padova, Italy}

\author{M. Aliotta}\affiliation{SUPA, School of Physics and Astronomy, University of Edinburgh, EH9 3FD Edinburgh, United Kingdom}

\author{H. Babu}\affiliation{Universit\`a degli Studi di Padova 35131 Padova, Italy}

\author{D. Bemmerer}\affiliation{Helmholtz-Zentrum Dresden-Rossendorf, 01328 Dresden, Germany}

\author{A. Boeltzig}\affiliation{Helmholtz-Zentrum Dresden-Rossendorf, 01328 Dresden, Germany}

\author{R. Depalo}\affiliation{Universit\`a degli Studi di Milano, 20133 Milano, Italy}\affiliation{INFN, Sezione di Milano, 20133 Milano, Italy}

\author{A. Caciolli}\affiliation{Universit\`a degli Studi di Padova, 35131 Padova, Italy} \affiliation{INFN, Sezione di Padova, 35131 Padova, Italy}

\author{F. Cavanna}\affiliation{INFN, Sezione di Torino, 10125 Torino, Italy}

\author{L. Csedreki}\affiliation{Institute for Nuclear Research (ATOMKI), PO Box 51, 4001 Debrecen, Hungary}

\author{Z. F\"ul\"op}\affiliation{Institute for Nuclear Research (ATOMKI), PO Box 51, 4001 Debrecen, Hungary}
\author{G. Imbriani}\affiliation{Universit\`a degli Studi di Napoli “Federico II”, 80126 Napoli, Italy}\affiliation{INFN, Sezione di Napoli, 80126 Napoli, Italy}

\author{D. Rapagnani}\affiliation{Universit\`a degli Studi di Napoli “Federico II”, 80126 Napoli, Italy}\affiliation{INFN, Sezione di Napoli, 80126 Napoli, Italy}

\author{S. Rümmler}\affiliation{Helmholtz-Zentrum Dresden-Rossendorf, 01328 Dresden, Germany}\affiliation{Technische Universität Dresden, Zellescher Weg 19, 01069 Dresden, Germany}

\author{K. Schmidt}\affiliation{Helmholtz-Zentrum Dresden-Rossendorf, 01328 Dresden, Germany}

\author{R.~S. Sidhu}\affiliation{SUPA, School of Physics and Astronomy, University of Edinburgh, EH9 3FD Edinburgh, United Kingdom}

\author{T. Sz\"ucs}\affiliation{Institute for Nuclear Research (ATOMKI), PO Box 51, 4001 Debrecen, Hungary}

\author{S. Turkat}\affiliation{Technische Universität Dresden, Zellescher Weg 19, 01069 Dresden, Germany}

\author{A. Yadav}\affiliation{Helmholtz-Zentrum Dresden-Rossendorf, 01328 Dresden, Germany}\affiliation{Technische Universität Dresden, Zellescher Weg 19, 01069 Dresden, Germany}

\begin{abstract}
The $^{12}$C(p,$\gamma$)$^{13}$N reaction is the onset process of both the CNO and Hot CNO cycles that drive massive star, Red and Asymptotic Giant Branch star and novae nucleosynthesis. The $^{12}$C(p,$\gamma$)$^{13}$N rate affects the final abundances of the stable $^{12,13}$C nuclides, with ramifications for meteoritic carbon isotopic abundances and the s-process neutron source strength.
Here, a new underground measurement of the $^{12}$C(p,$\gamma$)$^{13}$N cross-section is reported.
The present data, obtained at the Felsenkeller shallow-underground laboratory in Dresden (Germany), encompass the 320-620 keV center of mass energy range 
to include the wide and poorly constrained E = 422 keV resonance that dominates the rate at high temperatures. 
This work $S$-factor results, lower than literature by 25\%, are included in a new comprehensive R-matrix fit, and the energy of the $\frac{1}{2}^+$ first excited state of $^{13}$N is found to be 2369.6(4)~keV, with radiative and proton width of $0.49(3)$~eV and $34.9(2)$~keV respectively. A new reaction rate, based on present R-matrix fit and extrapolation, is suggested. 
\end{abstract}

\maketitle


\paragraph{Introduction $-$}
The $^{12}$C(p,$\gamma$)$^{13}$N reaction ($Q$-value = 1943.5(3)~keV \cite{Wang21-CPC}) is the first in the CNO cycle:
\begin{equation}
\begin{split}
 &^{12}{\rm C}(p,\gamma)^{13}{\rm N}(\beta ^{+}\,\nu)^{13}{\rm C}(p,\gamma)^{14}{\rm N}\xrightarrow{}\\
 &\xrightarrow{}^{14}\!\!{\rm
 N}(p,\gamma)^{15}{\rm O}(\beta ^{+}\,\nu)^{15}{\rm N}(p,\alpha)^{12}{\rm C} , 
\end{split}
\end{equation}
with the $^{13}$N $\beta^{+}$ decay ($t_{\rm 1/2}\,=\,9.965(4)$~min) being one of the sources of CNO neutrinos in the Sun, whose first direct observation was recently reported in \cite{Agostini20-Nature}.

H-burning via CNO cycle, at $T\,=\,$0.02 - 0.1~GK \cite{Iliadis07-Book}, is the main nuclear energy source in massive stars during main sequence phase and in stars in more advanced stages, \textit{i.e.} Red Giant Branch (RGB) and Asymptotic Giant Branch (AGB) stars \cite{Rolfs88-Book}.

In low-mass AGB stars the $^{12}$C(p,$\gamma$)$^{13}$N($\beta ^{+}\,\nu$)$^{13}$C
process is responsible for the so-called $^{13}$C pocket creation, which, through the $^{13}$C($\alpha$,n)$^{16}$O reaction, provides the main neutron source for $s$-process nucleosynthesis \cite{Straniero06-NPA, Gallino98-APJ}.
Recently, the $^{13}$C($\alpha$,n)$^{16}$O reaction cross-section was measured inside the $s$-process Gamow peak by both the LUNA and JUNA collaborations with reduction of extrapolation uncertainties \cite{Ciani21-PRL, Gao22-PRL}. The poorly constrained $^{12}$C(p,$\gamma$)$^{13}$N reaction rate and the large uncertainty on the mixing phenomena taking place in low-mass AGB stars, however, still have considerable impact on present predictions for the $^{13}$C-pocket formation and the subsequent $s$-process \cite{Arcones23-AAR}.

Moreover, the $^{12}$C(p,$\gamma$)$^{13}$N reaction directly affects the $^{12}$C/$^{13}$C isotopic ratio observed in presolar grains \cite{Zinner14-Chapter}, in the interstellar medium \cite{Milam05-APJ} and in the Solar system \cite{Scott06-AA}. Indeed the $^{12}$C/$^{13}$C ratio is a powerful tool to constrain nucleosynthesis and mixing processes in RGB and AGB stars, to explain the role of these stars as well as novae explosions in the galactic chemical evolution \cite{Milam05-APJ}.

At typical temperatures of nova explosion, $T\,\ge\,$0.1~GK \cite{Wiescher10-ARNPS}, the proton captures on $^{13}$N overtakes its $\beta^{+}$ decay leading to a different CNO cycle.
This variant, referred to as Hot CNO (HCNO), implies an enhanced nuclear energy generation and a major fraction of CNO nuclei to be transformed into $^{14}$O and $^{15}$O, whose decay energy shapes the nova light curve \cite{Ness03-APJL}. The temperature and density conditions of the transition from the CNO to the HCNO, as well as the nucleosynthesis output, strongly depend on reaction rates in the cycle as suggested in \cite{Mathews84-APJ}.
At higher temperatures, $T > 0.4$~GK and at critical values of the density, the HCNO is dominated by breakout reactions, mainly $\alpha$-capture on $^{14,15}$O nuclei, leading to the rapid proton capture ($rp$) process and triggering the conditions for X-ray bursts \cite{Wiescher10-ARNPS}.

In the energy range of astrophysical interest, up to 400~keV, the $^{12}$C(p,$\gamma$)$^{13}$N reaction cross-section is dominated by a poorly constrained broad resonance. The two most comprehensive studies available to date provide, indeed, conflicting results for the resonance energy \cite{Vogl63-PhD, Rolfs74-NPA}. In \cite{Rolfs74-NPA} the resonance energy was estimated in the laboratory frame at $E$\textsubscript{p}\footnote{In the following $E$\textsubscript{p} and $E$ will denote the proton energy in the laboratory and in the center of mass frame, respectively}$\,=\,$457(1)~keV, while \cite{Vogl63-PhD} found 462~keV, pointing out that a fit made with the resonance located at 456 keV, as previously reported in \cite{Jackson53-PR}, falls outside the range of uncertainty. The subsequent results by \cite{Burtebaev08-PRC} do not solve the discrepancy because of the few data points reported. In a recent work by \cite{Artemov22-EPJA} the $^{12}$C(p,$\gamma$)$^{13}$N reaction $S$-factor data were re-analysed in the light of a new determination of its asymptotic normalization coefficient. In this work the resonance is reported at $E\,=\,$425.3~keV, 3.6~keV higher than the adopted value \cite{Ajzenberg-Selove91-NPA}.
Crucial parameters for the transition from normal to HCNO are the poorly constrained resonance proton and radiative widths. The reported values, in the center of mass frame, for the former lay between $\Gamma$\textsubscript{p}$\,=\,$32-36~keV while the radiative widths available in the literature stay in the range of $\Gamma_{\gamma}\,=\,$0.50-0.63~eV  \cite{Rolfs74-NPA, Vogl63-PhD, Ajzenberg-Selove91-NPA, Artemov22-EPJA}, see Tab.\ref{table2}. 
Finally the $^{12}$C(p,$\gamma$)$^{13}$N reaction cross section was recently measured at ATOMKI via activation technique in a wide energy range \cite{Gyury23-EPJA}. No new parameters are determined for the resonance of interest here. The cross section results reported in \cite{Gyury23-EPJA} for the region of interest considered here are in good agreement with previous results from \cite{Rolfs74-NPA, Vogl63-PhD, Burtebaev08-PRC}.
Ultimately both, the resonance energy and total width, are crucial for the extrapolation down to stellar energies and thus for the reaction rate estimation. Extrapolation down to low energies, indeed, is not properly constrained by data reported for $E\,\le\,$320$\,$keV.
Four main data sets are available: the first two measurements were performed by the activation technique \cite{Bailey50-PR, Lamb57-PR}. The subsequent wide energy range measurements, by prompt $\gamma$ detection, are the already mentioned works by \cite{Vogl63-PhD} and by \cite{Rolfs74-NPA}.
All these data sets reported large uncertainties, up to 10\%, with data points scattering by 30\%. Recently a new comprehensive measurement was performed at Laboratory for Underground Nuclear Astrophysics, LUNA, employing different techniques and covering the $E\,=\,60-370\,$keV energy range \cite{skowronski-2023, Skowronski23-JPG}.

The aim of this work is to present a re-investigation of the $^{12}$C(p,$\gamma$)$^{13}$N reaction $S$-factor in the energy range $E$ = 320 - 620~keV. The experiment was performed at the shallow underground accelerator facility at Felsenkeller, Dresden (Germany) \cite{Bemmerer18-SNC}.

\begin{figure}[ht!]
\centering
\includegraphics[width=0.5\textwidth]{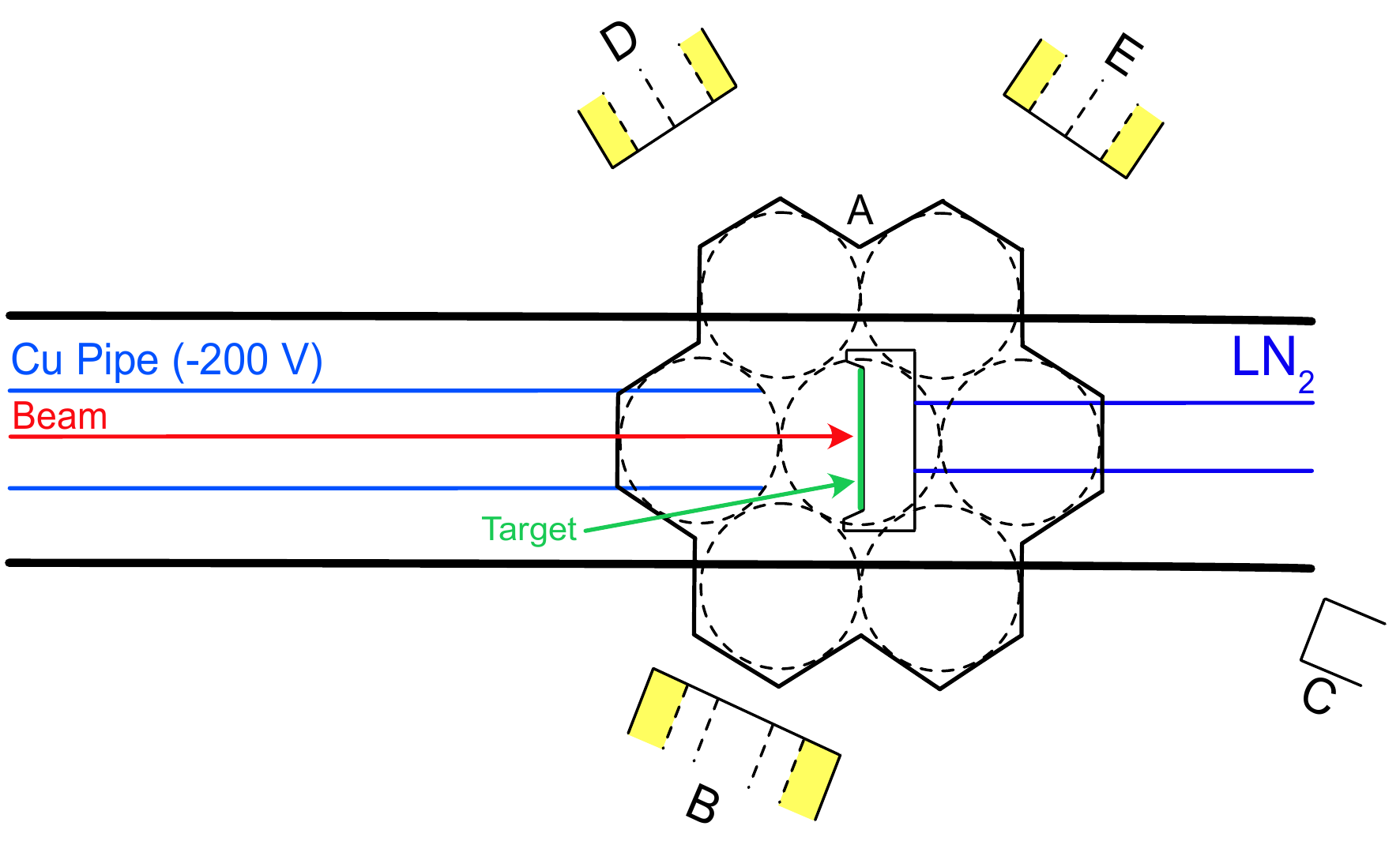}
\caption{Schematic top view of the present setup. The detector main features are summarized in Tab.\ref{table1}. Not to scale.}
\label{figure1}
\end{figure}

\paragraph{Experimental setup and data acquisition $-$}
 A schematic view of the experimental setup is shown in Fig. \ref{figure1}. The 5$\,$MV Pelletron accelerator of Felsenkeller Laboratory provided molecular H\textsubscript{2}$^{+}$ beam with proton energies ranging between $E$\textsubscript{p} = 350 and 670 keV. The beam energy of the accelerator was calibrated using the narrow $E$\textsubscript{p} = 991.86$\pm$0.03 keV \cite{2013NDS...114.1189S} resonance in the $^{27}$Al(p,$\gamma$)$^{28}$Si reaction \cite{2017NIMPB.406..108P}. For the calibration, this resonance was scanned using 0.5\,kV steps in nominal terminal voltage,  resulting in a proton energy determination with 0.5 keV uncertainty, corresponding to $\Delta k/k$ = 0.1\%, being $k$ the calibration parameter. 
This highly precise calibration point was confirmed in several ways: First, with the $^{14}$N($\alpha,\gamma$)$^{18}$F resonance doublet at 1527-1529 keV (0.4\% uncertainty). Second, using $^4$He$^+$ beams of 1.3-2.9 MeV beam energy with the analyzing magnet and with emitted direct-capture $\gamma$ rays (0.8\% uncertainty). Third, using N$_2^+$ and O$_2^+$ beams at 1.3-2.9 MV terminal voltage and the calibrated magnetic field (again 0.8\% uncertainty). Finally, using the measured current and measured resistivity along the high and low energy accelerator columns (3\% uncertainty). 

The beam was analyzed by a magnetic analyzer and collimated by three apertures and delivered through a copper tube, 65~mm long and with diameter of 22~mm. 
The Cu tube was positioned at 20 mm from the target target and it was in thermal contact with LN\textsubscript{2} to improve local vacuum conditions and prevent carbon build-up on the target.  The Cu tube was biased on a negative voltage of $-200$~V, for secondary electron suppression. To achieve a complete suppression of secondary electrons a permanent cubic magnet, 1~cm size, was added just below the target location, in a position that maximized electron suppression, according to dedicated tests.  

The target was mounted perpendicular to the beam direction on a copper holder in thermal contact with liquid nitrogen to limit target degradation. Both the target and the target chamber were electrically insulated from the beam line and served as a Faraday cup for beam current measurements. Throughout the experiment a typical H\textsubscript{2}$^{+}$ beam current of 10 $\mu$A, corresponding to 1.25$\times 10^{14}$ protons/s, was delivered on the target.

Two targets, namely L1 and L4, were produced at ATOMKI by evaporation of natural carbon powder (99.8\% nominal purity from ADVENT) on Ta disks (0.25 mm thick and with diameter of 27 mm from Goodfellow) previously cleaned both mechanically and chemically \cite{Ciani20-EPJA}.
Targets L1 and L4 were produced with slightly different thickness, 570 $\mathrm{\mathring{A}}$ and 550 $\mathrm{\mathring{A}}$ respectively, monitored during evaporation by a quartz installed in the vacuum evaporator. In addition, a natural graphite disk, 1~mm thick with 99.8\% nominal purity from ADVENT, was irradiated in the energy range $E$\textsubscript{p} = 380 - 450 keV for consistency check. 

The degradation of evaporated targets under beam irradiation was monitored in-situ via two independent techniques: the peak-shape analysis and the Nuclear Resonant Reaction Analysis (NRRA)~\cite{Ciani20-EPJA}. Periodic runs were performed at $E$\textsubscript{p} = 380 keV and the $^{12}$C(p,$\gamma$)$^{13}$N $\gamma$-peak was analysed. For NRRA measurements the well known $E$\textsubscript{p} = 1747.6(9) ~keV narrow resonance ($\omega\gamma$ = 7.3(5) eV, $\Gamma$ = 135(8)~eV \cite{Zeps95_PRC}) in the $^{13}$C(p,$\gamma$)$^{14}$N reaction was exploited.
Scans of the resonance, with $E$\textsubscript{p} = 1745 - 1756 keV, were performed on both evaporated targets at the beginning and at the end of the data acquisition, at accumulated charge of about 7 and 2 ~C, respectively. A comparison with a similar investigation performed in ATOMKI soon after target production showed a good agreement. Both, peak-shape analysis and the resonance scans, did not reveal any target degradation during the measurement.

In the energy range investigated here the $^{12}$C(p,$\gamma$)$^{13}$N reaction emits a single $\gamma$-ray, $E_{\gamma}$ = $Q$-value + $E$, which was detected with a 7-HPGe crystals cluster detector, labeled A in Tab.\ref{table1} and Fig.\ref{figure1}, positioned at 90(2)$^\circ$ with respect to the beam direction and at 6.1(2)~cm from the center of the target. The $\gamma$-ray angular distribution was reported to be isotropic in \cite{Rolfs74-NPA, Burtebaev08-PRC} at energies of interest here, consistent with $J^{\pi}\,=\,\frac{1}{2}^+$ assignment of $E$\textsubscript{x} = 2364.9(6)~keV excited state \cite{Wang21-CPC}. 
At three proton energies, namely $E$\textsubscript{p} = 400, 464 and 555~keV, the angular distribution was checked using four detectors placed at different angles and distances all around the target chamber, see Tab.\ref{table1} and Fig.\ref{figure1} for details.

The absolute full-energy peak efficiency of all detector crystals was measured using point-like radioactive sources ($^{137}$Cs, $^{60}$Co and $^{22}$Na), with activities calibrated by the Physikalisch-Technische Bundesanstalt (PTB) to 1\% accuracy. The efficiency curve was extended up to 10 MeV using the well-known $^{27}$Al(\textit{p},$\gamma$)$^{28}$Si resonance at proton energy $E$\textsubscript{p} = 992~keV \cite{
Harissopulos00-EPJA}. For detector A the full-energy efficiency, corrected for the true coincidence summing effect (3\%), at the $\gamma$-ray energy of interest for the $^{12}$C(p,$\gamma$)$^{13}$N reaction, $\xi_\gamma$, was obtained through the analytic fit of experimental data as described in \cite{Imbriani05-EPJ}. 
The summing effect was negligible for the detectors used for determining the angular distribution, because of their large distance from the target.
The efficiency uncertainty was estimated as the sum of different contributions. The uncertainty from the fit itself was estimated as the average residual between the measured and the calculated yields. For the summing correction a conservative uncertainty of 50\% was assumed, given the small impact on the efficiency results.
Finally the effect on efficiency of the observed beamspot position was estimated to be 5\% by calculating the covered solid angle in case of a centered beam and a 5~mm off-centered spot. The total uncertainty of the efficiency is 6.5\%.

\begin{table}[]
    \centering
    \caption{Detectors used, their types, size, angle to the target normal (uncertainty 2~deg), and distance from detector endcap to target center (uncertainty 0.2 cm).}
    \label{table1}
    \begin{tabular}{c c c c c}
    \midrule
    \midrule
        Detector ID & Cluster Type & Relative Efficiency & Angle & Distance \\
        & &{\%} & [deg] & [cm] \\
        \midrule
        A &  7 crystals & 7$\times$60 & 90 & 6.1 \\
        B & 7 crystals & 7$\times$60 & 114 & 19.2 \\
        C & single crystal & 100 & 22 & 28.7 \\
        D & 3 crystals & 3$\times$60 & 122 & 44.1\\
        E & 3 crystals & 3$\times$60 & 55 & 42.8 \\
        \midrule
        \midrule
    \end{tabular}
\end{table}

\paragraph{Data analysis $-$}
The acquisition was in list mode for all crystals, namely for each event both energy and time were recorded, and two additional channels were dedicated to the acquisition chain for the beam current on the Cu tube and on the target. Two different data acquisition (DAQ) boards were used one with high gain, used to study the $^{12}$C(p,$\gamma$)$^{13}$N reaction signal, and one with low gain, used to detect high energy $\gamma$-rays from the $^{13}$C(p,$\gamma$)$^{14}$N reaction during the target scans and at the same time to have a fine binning for the peak-shape analysis of the $^{12}$C(p,$\gamma$)$^{13}$N reaction $\gamma$-ray peak.

The experimental yield, $Y$, for the $^{12}$C(p,$\gamma$)$^{13}$N peak was obtained for each run as follows:

\begin{equation}
Y = \frac{N_{\gamma}}{N_p \cdot \xi_{\gamma} \cdot W(\theta)}
\label{eq_1}
\end{equation} 

\begin{figure*}[ht!]
\includegraphics[width=0.9\textwidth]{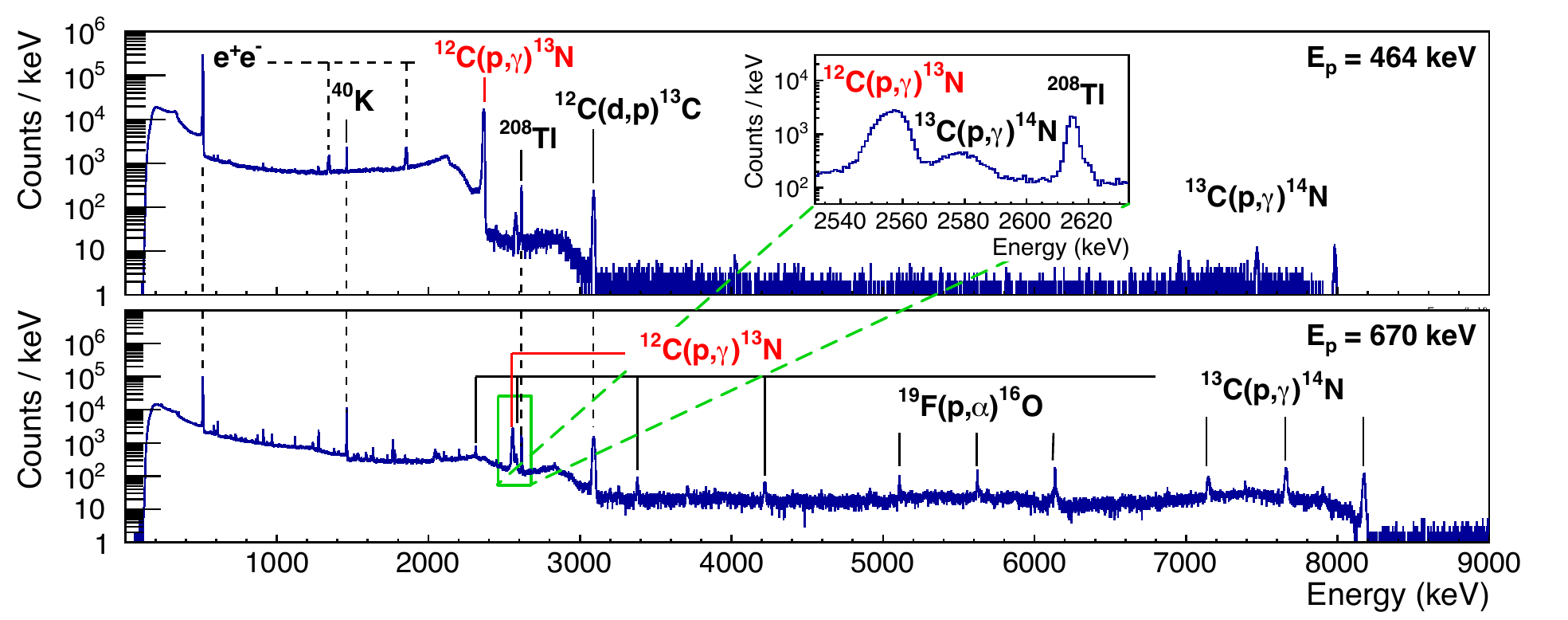}
\caption{The $\gamma$-ray spectra acquired at $E$\textsubscript{p} = 464~keV, top, and at 670~keV, bottom. The $\gamma$-ray from the $^{12}$C(p,$\gamma$)$^{13}$N reaction is labelled in red. The typical environmental and beam induced background $\gamma$-rays are indicated.}
\label{figure2}
\end{figure*}

\begin{figure}[ht]
\includegraphics[width=0.5\textwidth]{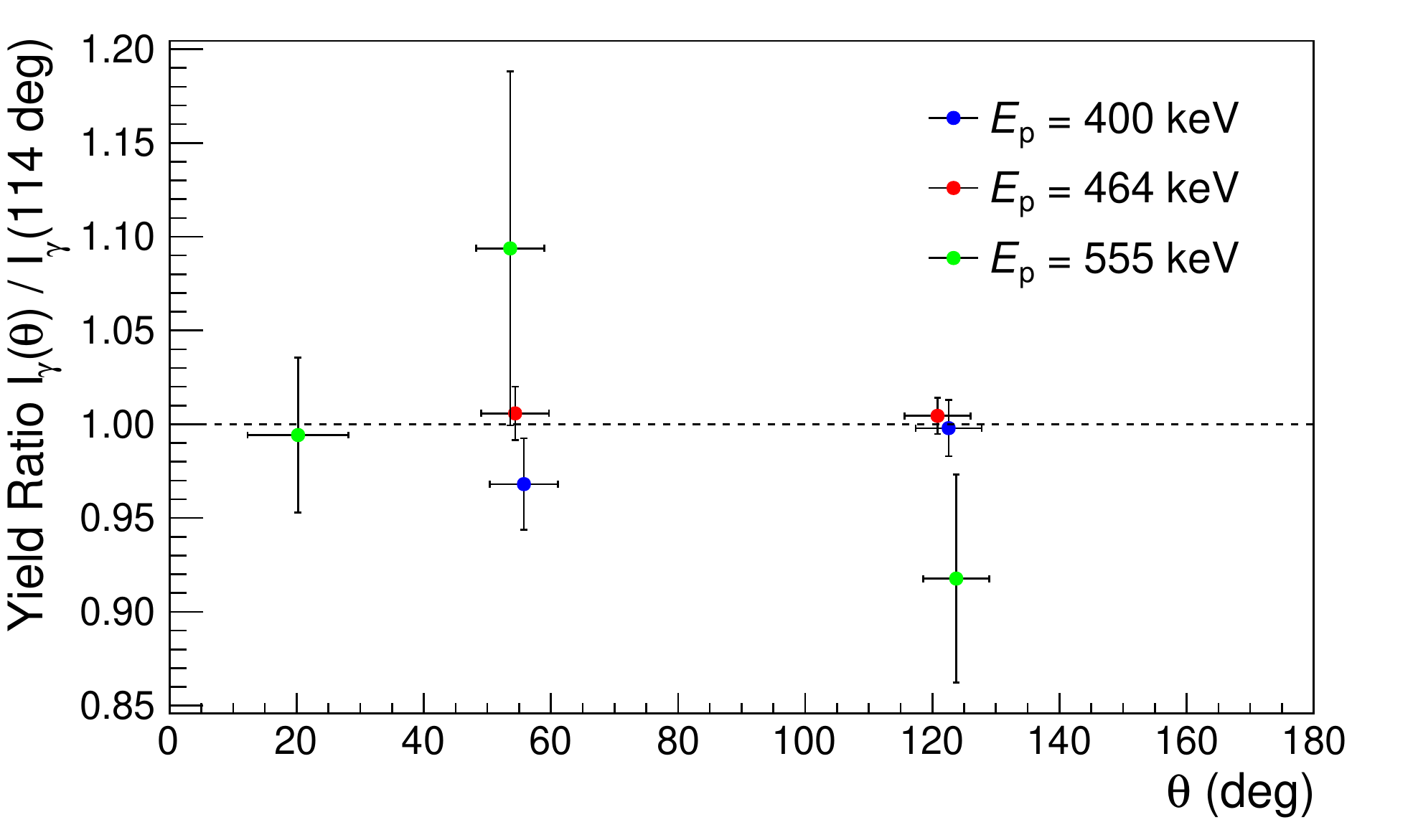}
\caption{Comparison between yields from detectors B (114~deg), C (22~deg), D (122~deg) and E (55~deg) at three different beam energies. The data confirm that the $\gamma$-ray of interest is isotropic as expected. Detector C was removed for technical reasons and replaced in a different position before runs at $E$\textsubscript{p} = 400 and 465~keV without re-performing a proper efficiency calibration, thus these data are not included. The x-axis error bars include the contribution from the detector position uncertainty (2~deg) and the opening angle of the detector (8~deg for C and 5~deg for detectors B and D).}
\label{figure3}
\end{figure}

The net counts $N_{\gamma}$ were obtained for each crystal of HPGe cluster A with typical statistical uncertainty of 1-2\%. The number of impinging protons, $N_p$, was derived from the accumulated charge. The uncertainty of the charge collection is conservatively estimated at 3\% level. A 0.07(1)\% deuterium contamination in the molecular beam was estimated by the analysis of the $^{12}$C(d,p)$^{13}$C reaction peak at 3090~keV \cite{Putt71, Debras77}, see Fig.\ref{figure2}. 
Nevertheless, its contribution to the total charge is negligible. The efficiency at the $E_{\gamma}$ of interest is $\xi_{\gamma}$.
The angular distribution coefficient $W(\theta)$ is 1 as expected and confirmed, within 5\%, from results of the dedicated runs performed with detectors B, C, D and E, see Fig.\ref{figure3}.

For the calculation of the astrophysical $S$-factor, $S(E)$, we used the following relationship \cite{Rolfs88-Book}:

\begin{equation}
Y = \int {E^{-1} S(E) e^{-2\pi\eta(E)} \epsilon_{\mathrm{eff}}^{-1}(E)P(E) dE}
\label{eq_1}
\end{equation} 

where $E$ is the proton beam energy in the center of mass frame and $\eta(E)$ is the Sommerfeld parameter \cite{Rolfs88-Book}.  The effective stopping power in the laboratory frame, $\epsilon_{\mathrm{eff}}(E)$, was calculated for the natural isotopic composition of carbon (99\% $^{12}$C and 1\% $^{13}$C) using SRIM2013 database \cite{Ziegler10-NIMPRB}. Target stoichiometry was checked throughout the experiment using the runs on top of the 1.7~MeV resonance of the $^{13}$C(p,$\gamma$)$^{14}$N reaction. The results were always consistent with the natural isotopic abundances. The effective stopping power uncertainty is of 3.5\% as follows from experimental values reported in
SRIM in the energy range of interest here.

The target profile $P(E)$ was obtained from NRRA results, properly corrected at each beam energy for the different energy loss and straggling. The profile was parameterized as reported in \cite{Ciani20-EPJA}, and target thicknesses of 13.3(4) and 15.8(4)~keV at $E$\textsubscript{p} = 380 keV were found for L1 and L4, respectively. The results for the $S(E)$ was proved to be weakly dependant on the target profile and $S$-factor energy dependency assumed.

Finally, the energy associated to the resultant $S(E)$ is the effective energy, as defined in \cite{Brune13-NIMA}.
\paragraph{Results and discussion $-$}
The new $^{12}$C(p,$\gamma$)$^{13}$N reaction $S(E)$ results are shown in Fig.\ref{figure4}. See Supplemental Material \cite{SM} for a table with $S$-factor values.

\begin{figure}[ht]
\includegraphics[width=0.5\textwidth]{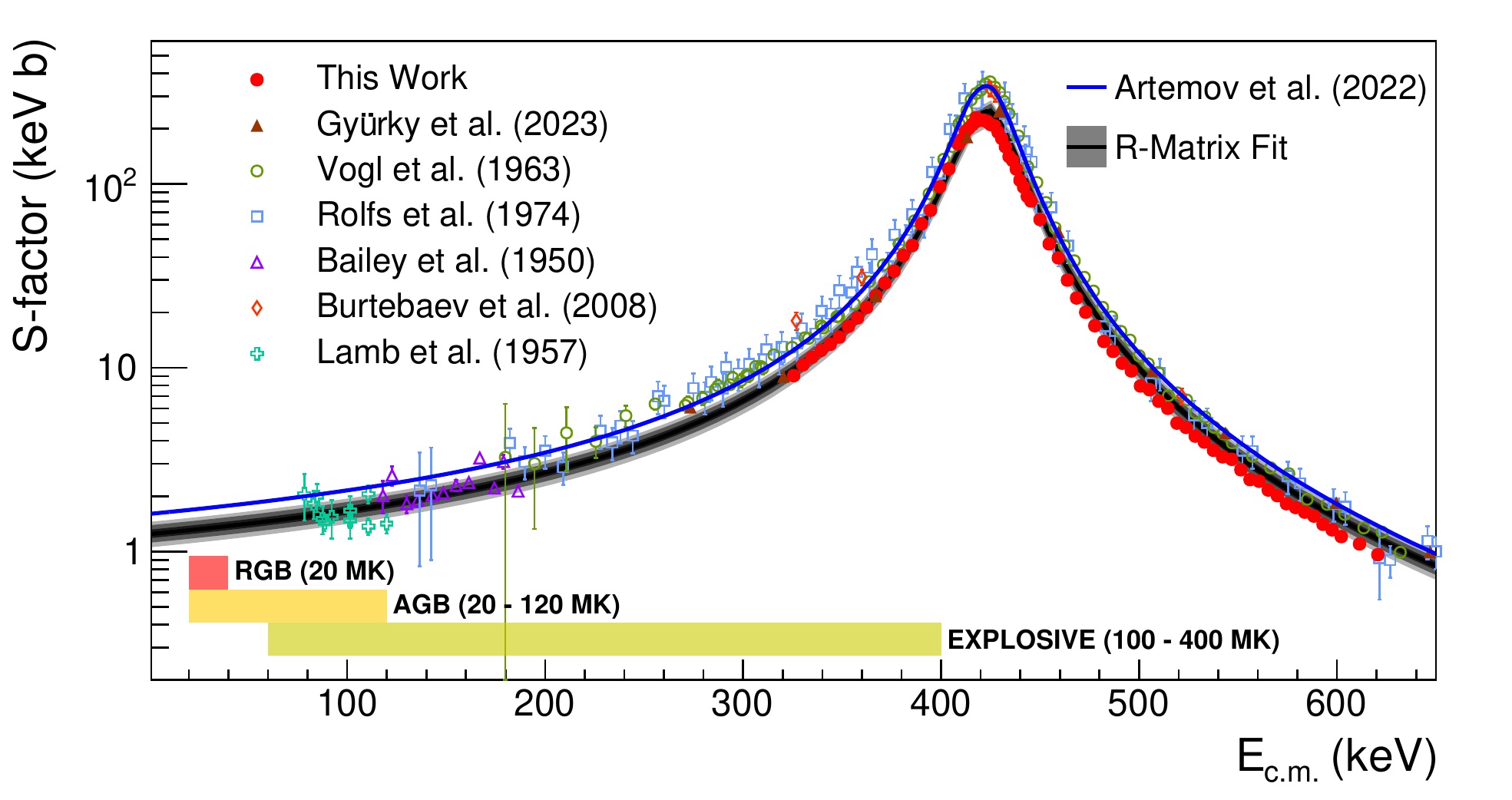}
\caption{Present results for the $^{12}$C(p,$\gamma$)$^{13}$N reaction $S(E)$, red points, compared with data available in literature. The present R-matrix fit is shown in black and compared with the similar fit reported in \cite{Artemov22-EPJA}, blue line.}
\label{figure4}
\end{figure}

Only statistical uncertainties are plotted in Fig.\ref{figure4}. The systematic uncertainty of $S(E)$ amounts to 8.5\% including the contributions from detection efficiency (6.5\%), stopping power (3.5\%), target profile (3\%) and charge collection (3\%).

Over the whole energy range a 25\% scaling difference is observed between present results and literature data available in the same energy range \cite{Vogl63-PhD, Rolfs74-NPA,Burtebaev08-PRC, Gyury23-EPJA}.

An R-matrix analysis was performed using AZURE2 code \cite{Azuma10-PRC} and considering all the data available with two-fold aim, get new resonance parameters and extrapolation down to energies of astrophysical interest. 

The included channels and the parameters were taken from \cite{Azuma10-PRC}, alongside the channel radius of $3.4$ fm. The asymptotic normalization constant (ANC) value was updated to the value recently reported in \cite{Artemov22-EPJA}. All the datasets reported in Fig. \ref{figure4} were considered in the fit. 
Additionally, the proton scattering data from \cite{Meyer76} were used to constrain the elastic channel as well. For the studies that do not report any systematic uncertainty, a value of 20\% was considered. The present results are included with the error budget aforementioned and 
The minimization and the error estimation was handled through the bayesian approach using the BRICK framework \cite{Odell22}. Only uninformative priors were used apart from the ANC, which distribution was assumed to be gaussian. Additionally, the normalization factors were treated as free parameters with prior distributions defined as gaussians peaked at 1 and sigma given by their systematic error. Apart from the systematic and statistical errors, the energy calibration error was included as well in the fitting procedure \cite{Odell22}.

The result of the R-matrix fit is shown in black in Fig.\ref{figure4}. It is mainly constrained by the present work, due to the high density of data and the reduced uncertainty. The extrapolated $S(25)$ from the present fit is $1.34(9)$~keV b, lower by $23$\% than the result recently reported in \cite{Artemov22-EPJA}.

The new best-fit resonance parameters are shown in Tab.\ref{table2}, together with literature values. Our results for the resonance energy is in good agreement with values reported in \cite{Vogl63-PhD} but with improved precision, while a significant discrepancy is observed with respect to the previous experimental works \cite{Burtebaev08-PRC, Rolfs74-NPA}. 

The present resonance radiative width is 22\% lower than the most recent analysis by \cite{Artemov22-EPJA, Burtebaev08-PRC}, while a good agreement is found when compared with the older data adopted in \cite{Ajzenberg-Selove91-NPA}. On the other hand our proton width is in good agreement with \cite{Burtebaev08-PRC, Rolfs74-NPA} while it is not consistent with values reported in \cite{Vogl63-PhD, Ajzenberg-Selove91-NPA}.

\begin{table}[]
    \centering
     \caption{Resonance parameters in the center of mass for the excited state at 2369.6(4) keV from present work (see text for details) compared to literature data with corresponding uncertainty if available.}
    \begin{tabular}{l l l c c}
    \midrule
    \midrule
    $E$ [keV]& $\Gamma_{\gamma}$ [eV] & $\Gamma$\textsubscript{p} [keV] 
    & Reference\\
        \midrule
        $426.1(4)$ & $0.48(3)$  & $35.6(2)$ & This Work &\\  
            $425.3$ & $0.63(7)$ & $33.5(10)$ & \citet{Artemov22-EPJA}&\\
                $421$ & $0.65(7)$ & $35(1)$ & \citet{Burtebaev08-PRC}\\
               421.7(5) & 0.50(4) & 31.7(8) & \citet{Ajzenberg-Selove91-NPA}\\
               $421.6(10)$ & & $36(2)$ & \citet{Rolfs74-NPA}&\\
               $426.2$ &  & 33 & \citet{Vogl63-PhD}&\\
        \midrule
        \midrule
    \end{tabular}
    \label{table2}
\end{table}

\begin{figure}[!h]
\includegraphics[width=0.5\textwidth]{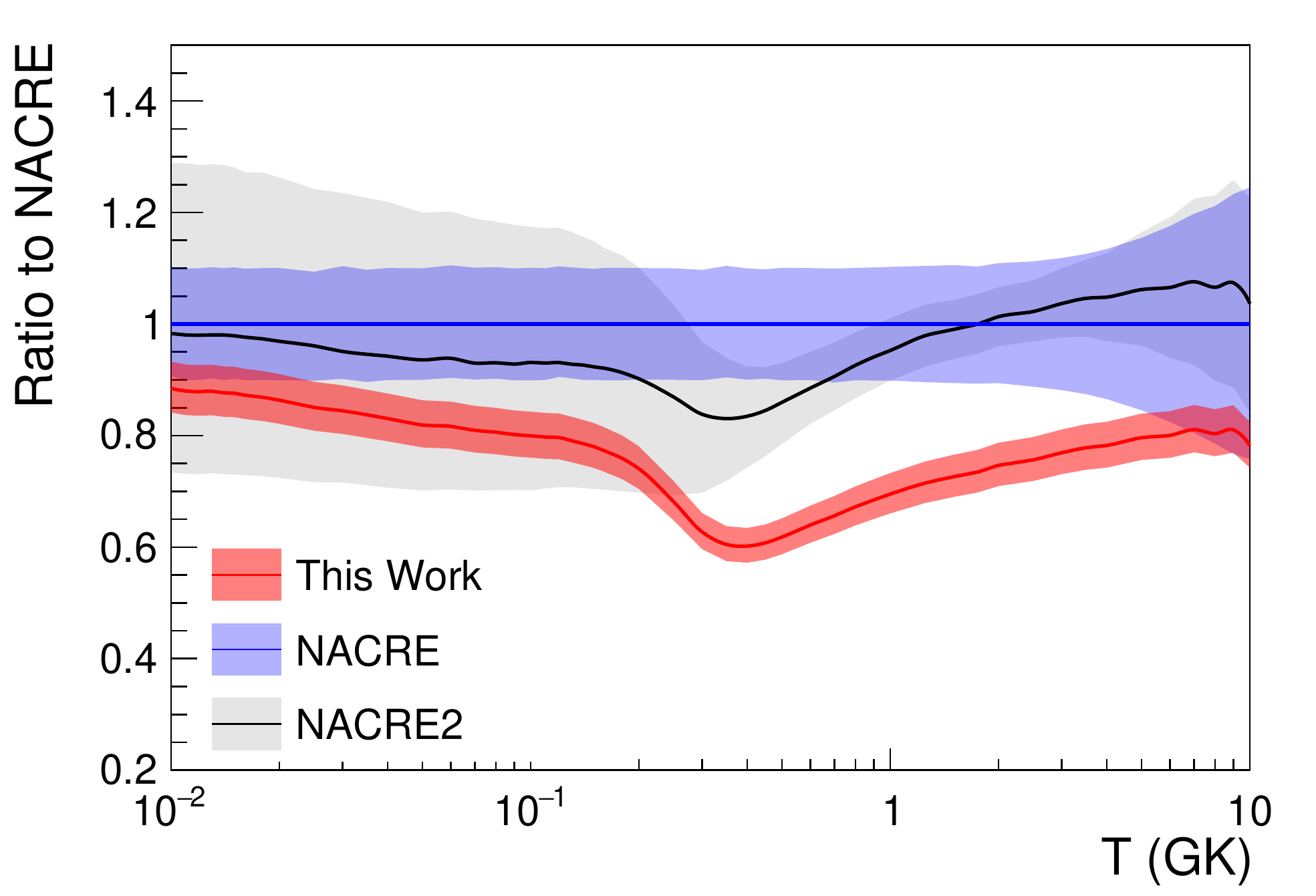}
\caption{The astrophysical reaction rate from the present work (red), normalized to NACRE \cite{Angulo99-NucPhysA}. The uncertainty of the present rate is within $9$\% over the entire temperature range. The reaction rate in \cite{Xu13-NucPhysA}, normalized to NACRE, is reported for comparison.}
\label{figure5}
\end{figure}

To evaluate the impact of our results, the thermonuclear reaction rate was calculated with the present R-matrix fit of the $S$-factor, see Supplemental Material \cite{SM}, and compared with the most widely adopted reaction rates \cite{Angulo99-NucPhysA,Xu13-NucPhysA}, see Fig.\ref{figure5}.

The present reaction rate uncertainty is significantly reduced compared to both \cite{Angulo99-NucPhysA,Xu13-NucPhysA} over the whole temperature range, 0.01-10~GK. In particular, at $T\,>\,0.3$~GK the present reaction rate falls outside \cite{Angulo99-NucPhysA,Xu13-NucPhysA} uncertainty, being 20\%-40\% lower.

\paragraph{Conclusion $-$}
The $^{12}$C(p,$\gamma$)$^{13}$N reaction cross-section has been measured in a wide energy range, 350$<E$\textsubscript{p}$<$670~keV, at the Felsenkeller facility, with a total uncertainty of about 9\%.
The present $S(E)$ results show a discrepancy of about 25\% with respect to data available in the literature over the whole energy range explored, requiring further experimental investigations, particularly at low energies.
A comprehensive R-matrix analysis has been performed to derive new precise parameters for the resonance, which are found to be in agreement with \cite{Vogl63-PhD}. Stringent new values for the radiative and proton widths were also derived.
Finally, the calculated reaction rate is consistently lower than literature, particularly at temperatures higher than $0.3$ GK suggesting a revision of the stellar model calculations for explosive H-burning and the need for a renewed evaluation of the impact on HCNO nucleosynthesis.

\paragraph{Acknowledgments $-$}
The authors would like to thank
all the Felsenkeller staff members for their technical support. Gy. Gy\"urky from ATOMKI for the help in producing targets.
Financial support by INFN, the
Italian Ministry of Education, University and Research (MIUR) through the ``Dipartimenti di eccellenza'' project ``Science of the Universe'', European Union, ChETEC-INFRA (grant agreement no. 101008324),  the Hungarian National Research, Development and
Innovation Office (NKFIH K134197, PD129060 and FK134845),  are gratefully acknowledged. M. A. acknowledges funding by STFC UK (grant no. ST/L005824/1). D.R. acknowledges funding from the European Research Council (ERC) under grant agreement no. 852016. T.S. acknowledges support from the J\'anos Bolyai research fellowship of the Hungarian Academy of Sciences and from the New National Excellence Programs of the Ministry of Human Capacities of Hungary under nr. \mbox{\'UNKP-22-5-DE-428}.

\end{document}